\documentclass[manuscript,screen]{acmart}

\AtBeginDocument{%
  \providecommand\BibTeX{{%
	\normalfont B\kern-0.5em{\scshape i\kern-0.25em b}\kern-0.8em\TeX}}}

\setcopyright{rightsretained}
\copyrightyear{2023}
\acmYear{2023}
\acmDOI{}

\acmConference[CCAI 2023]{CHI 2023 Workshop on Child-centred AI Design: Definition, Operation and Considerations}{April 23, 2023}{Hamburg, Germany}

\acmBooktitle{CHI 2023 Workshop on Child-centred AI Design: Definition, Operation and Considerations, April 23, 2023, Hamburg, Germany}

\begin{document}

\title{Towards Goldilocks Zone in Child-centered AI}


\author{Tahiya Chowdhury}
\affiliation{%
  \institution{Davis Institute for Artificial Intelligence}
  \city{Waterville, Maine}
  \country{USA}}
\email{tahiya.chowdhury@colby.edu}

\renewcommand{\shortauthors}{Chowdhury et al.}

\begin{abstract}

Using YouTube Kids as an example, in this work, we argue the need to understand a child's interaction process with AI and its broader implication on a child's emotional, social, and creative development. We present several design recommendations to create value-driven interaction in child-centric AI that can guide designing compelling, age-appropriate, beneficial AI experiences for children.

\end{abstract}



\maketitle

\section{YouTube Kids: A case of child-centered AI}
My 7-year-old nephew loves paper crafts and once asked for my help to make an origami frog. Sitting next to him, I created an account on YouTube Kids to find origami tutorials. As the homepage showed up with 28 videos neatly organized in a grid, I immediately found him interested in several videos, none of which is related to origami. 
Meanwhile, I noticed one video on the homepage containing adolescents wearing provocative clothes, which is age inappropriate for him according to his parents. All this happened without a single click made or a video watched on a platform designed for children.

As I am writing this, YouTube Kids has 35 million kids viewers every week \cite{youtube_blog}. Children's content creation on YouTube has become a lucrative business as evidenced by hundreds of YouTube videos available on the topic `How to Make Money on YouTube Kids'. The kids' contents are created by parents, kids, teenagers, families -- almost everyone. The top content categories watched on YouTube Kids are nursery rhymes and animated songs, created by a selected few channels with high viewership. The videos often include a child actor or cartoon character depicting a fantasy-filled life, with magical creatures, family, and friends doing fun-filled activities (interested readers can search for `Cocomelon' or `Like Nastya Show' on YouTube).

Parents find value in YouTube for both entertainment and educational purposes. 50\% of US parents of a child aged under four years let their kids watch YouTube videos daily primarily for entertainment purposes \cite{pew_research}. Half of these parents have found that their children have encountered inappropriate content on the app. Exposure to such content occurs through videos \cite{10.1145/3341161.3342913}, comments \cite{10.1145/3442442.3452314}, video language (and subtitle) and advertisements \cite{https://doi.org/10.48550/arxiv.2211.02356} presented to the user by the app's recommendation algorithm.

We are currently at a pivotal moment. For decades we have known that every child is unique, in their learning style and personality. With artificial intelligence-enabled applications and services, we now have an opportunity to cater to the unique needs of a child, a feat difficult to achieve through traditional classroom learning. Through connected devices, we can collect ubiquitous data about a child's behavior, identify their interests and struggles, and develop personalized curricula or therapy centered around nurturing the child's abilities. This offers an opportunity for early detection of child development issues such as autism spectrum disorder and to design personalized therapy robots to help the parents enable the child to become an active part of society \cite{Rudovic_2018}. In reality, our children's daily interaction with AI is largely dominated by apps such as YouTube.

A child's interaction with YouTube is purely one-sided, as YouTube disables comments on videos targeted towards child audience due to COPPA requirements \cite{coppa}. 
Note that much of the content available on the platform are auto-generated videos made by bots, often combining child-friendly characters with adult content involving violence and published with uncomfortable titles such as 'Wrong Heads Disney Wrong Ears Wrong Legs Kids Learn Colors Finger Family 2017 Nursery Rhymes'. Hundreds of videos can thus be uploaded per week (the per day upload limit for an account is 30 videos), and slip through YouTube's automatic content moderation mechanism. Prolonged screen exposure has several negative consequences such as disrupted sleep, poor communication skills, speech delay, etc. It is perhaps not a coincidence that autism rates have nearly tripled in the last decade, with an increased rate for symptoms such as impaired social skills and difficulty in communication \cite{autism}. 

YouTube is a powerful platform with immense potential. Despite the negative impacts on mental health, it is the only social media platform reported to have a positive impact on young individuals, providing them access to skill development and community building \cite{rsph}. To reap the potential of any child-AI interaction, we need to identify the values we seek from such interaction and embed these values into the design principles. 

To understand the nature of child-AI interaction, let us consider a child's information and entertainment-seeking process off (condition 1) and on YouTube (condition 2). 
Ayana, who is 12 years old interested to learn origami. In condition 1, Ayana goes to their local library to find 5-6 books on this topic. They found 2-3 books from this collection irrelevant to their interest after a quick exploration and then checks out the other 3 books. They encountered issues following the guidelines in the books and felt the need for someone to demonstrate the steps for them to follow. The overall process takes longer but remains focused on a single activity throughout the interaction.

In condition 2, Ayana uses YouTube to learn about origami. On their first visit to the app (in the presence of an adult guardian), videos ranked by high view counts are displayed on the homepage. 
According to YouTube, `there’s an audience for almost every video, and the job of our recommendation system is to find that audience' \cite{YouTube_rec}. The two key signals YouTube uses in its recommendation are: clicks and watch time. These signals are used to populate users' homepage with new videos ranked by the relevance to the clicks users made and the videos they watched. The homepage contents will be curated by activities of others in their demographic group, using `collaborative filtering' technique.



In condition 1, while the user spends more time finding the information they seek, they do not experience the impulse for task switching and overcome resistance to return to their original goal. In case they indeed explore other contents, that interaction remains independent, not influencing their future experience (e.g. it is unlikely that the author will find their regular study desk at the library filled with books on cats because they opened and peeked into a few pages of a title
`Knit your Own Cat` on their last visit). It is also highly improbable that the bookshelves around Ayana will be filled with new content on teenage singers as that is 
what people from similar demographic and locations are watching. The user thus has more control over their interaction and experience in condition 1. 

The form of social media interaction we see in condition 2  
has implications that can span many aspects of children's life. 
The children interacting with AI are among the first generation being surveilled right from their birth, through their or their parent's digital history. Worldwide, children start viewing YouTube content as young as 2 years old and gain repeated exposure to omni-available, user-generated, interaction-driven entertainment platforms from a very young age. At this age, they also begin their journey for emotional, social, creative, and physical growth, and slowly shape their view of how the world works.

\section{Towards value-driven interaction: A Goldilocks Zone}

A healthy child development process involves self-concept, curiosity, confidence, and motivation to learn and explore. And this process fosters an optimal learning environment where a child is free to make errors and improve. 
The consequences of being exposed to `what million others of their age are thinking' 
can impair their self-esteem development and their emerging interests can be dominated by thousands of others currently in trend. 
The lack of agency in AI-driven automated experience can exacerbate as their momentary wandering will influence their interaction experience and achieving the values sought from the platform will become difficult. 

We argue that this is a concern that needs attention from parents, policymakers, and designers of children-AI interaction. The concern stems from the lack of design principles for child-appropriate technology rather than issues found within a certain platform such as YouTube. At the heart of any recommendation system lies collaborative filtering, which finds content that other similar users are enjoying. Based on this, we point out that such a social recommendation system used in child-AI interaction is detrimental to a child's early cognitive development, social skills, self-concept, creativity, and curiosity. To that end, some recommendations for child-centered AI design are presented below:

\begin{itemize}
    \item Re-purposing adult-focused technology for children relies on a reactive mechanism for safeguarding and this has to change. On YouTube, content creators are responsible for the content and the user (parents in the case of a child user) for their activities (see YouTube's user agreement)\cite{YouTube_terms}. Between the responsibility of the creator (the sender) and the viewer (the receiver) lies the responsibility of the platform (the service provider), responsible for the recommendation algorithm selecting content for the viewer. Child-AI interaction should have a clear accountability principle for negative impacts resulting from AI algorithms.

    \item {Understanding the child-centered AI context requires involving both children and their parents (or guardians) in the design process. Parents are the primary caregiver responsible for child safety and happiness. As AI algorithms favor majority groups in the data, on culture and value conflict issues (contents on YouTube kids in 2022 largely reflect western cultures), incorporating parents can help achieve fairness and value-driven interaction.}

    \item{Child-centric AI experience should be free of collaborative recommendation which takes away control of interaction from the child. Design choices should be guided by values expected by the service and by age-appropriate consideration of the cognitive development they experience.}

    \item{Realizing the potential of child-centric AI will require large-scale, diverse data collected from child users. Though children's interactions and responses are different from adults, the lack of child data persists due to consent and privacy constraints.  As AI systems will be more pervasive in how children play, learn, and transition into adults, an appropriate framework needs to be formulated in collaboration with children, parents, designers, child psychologists, and policymakers for ethical data collection and user studies.} 
    
    
\end{itemize}

For designing child-centered AI, or any user-centered AI, it is vital to identify the core values to be embedded and prioritized in the interaction design. Child-centered AI research can benefit from developing a framework of 'Contextual Safeguarding' \cite{10547/624844} for researchers and practitioners working in this space. As a community focused on improving human interaction and creative processes through technology, we hope child-centered AI platforms such as YouTube Kids can provide practical lessons to inform the design of future experiences.


\bibliographystyle{ACM-Reference-Format}
\bibliography{sample-base}

\end{document}